\documentclass[aoas,MSNbibl,nameyear,dvips]{arximspdf}
\usepackage{multirow}
\usepackage{dcolumn}
\usepackage{graphicx}
\usepackage{breakurl}

\doi{10.1214/13-AOAS671} 
\volume{7}
\issue{4}
\pubyear{2013}
\firstpage{1898}
\lastpage{1916}

\makeatletter

\newcolumntype{d}[1]{D{.}{.}{#1}}

\newcommand{\blX}{\mathbf{X}}
\newcommand{\blY}{\mathbf{Y}}
\newcommand{\blD}{\mathbf{D}}

\makeatother

\begin{document}
\begin{frontmatter}

\title{An algorithm for deciding the number of clusters and validation using
simulated data with application to exploring crop population structure}
\runtitle{Deciding the number of clusters}

\begin{aug}
\author[A]{\fnms{Mark A.} \snm{Newell}\corref{}\thanksref{t1,m1}\ead[label=e1]{manewell@noble.org}},
\author[B]{\fnms{Dianne} \snm{Cook}\thanksref{t2,m2}\ead[label=e2]{dicook@iastate.edu}},
\author[B]{\fnms{Heike} \snm{Hofmann}\thanksref{t2,m2}\ead[label=e3]{hofmann@iastate.edu}}\\
\and
\author[C]{\fnms{Jean-Luc} \snm{Jannink}\thanksref{m3}\ead[label=e4]{jeanluc.jannink@ars.usda.gov}}
\runauthor{Newell, Cook, Hofmann and Jannink}
\affiliation{The Samuel Roberts Noble Foundation\thanksmark{m1},
Iowa State University\thanksmark{m2} and Cornell
University\thanksmark{m3}}
\address[A]{M. A. Newell\\
The Samuel Roberts Noble Foundation\\
2510 Sam Noble Parkway\\
Ardmore, Oklahoma 73401\\
USA\\
\printead{e1}} 
\address[B]{D. Cook\\
H. Hofmann\\
Department of Statistics\\
Iowa State University\\
2413 Snedecor\\
Ames, Iowa 50011\\
USA\\
\printead{e2}\\
\hphantom{E-mail: }\printead*{e3}}
\address[C]{J.-L. Jannink\\
USDA-ARS\\
Robert W. Holley Center\\
\quad for Agriculture and Health\\
Department of Plant Breeding and Genetics\\
Cornell University\\
Ithaca, New York 14853\\
USA\\
\printead{e4}}
\end{aug}

\thankstext{t1}{Supported by USDA-NIFA Grant 2008-55301-18746.}
\thankstext{t2}{Supported in part by NSF Grant DMS-07-06949.}

\received{\smonth{7} \syear{2011}}
\revised{\smonth{6} \syear{2013}}

%
\begin{abstract}
A first step in exploring population structure in crop plants and other
organisms is to define the number of subpopulations that exist for a
given data set. The genetic marker data sets being generated have
become increasingly large over time and commonly are of the
high-dimension, low sample size (HDLSS) situation. An algorithm for
deciding the number of clusters is proposed, and is validated on
simulated data sets varying in both the level of structure and the
number of clusters covering the range of variation observed
empirically. The algorithm was then tested on six empirical data sets
across three small grain species. The algorithm uses bootstrapping,
three methods of clustering, and defines the optimum number of clusters
based on a common criterion, the Hubert's gamma statistic. Validation
on simulated sets coupled with testing on empirical sets suggests that
the algorithm can be used for a wide variety of genetic data sets.
\end{abstract}

%

\end{frontmatter}

\section{Introduction}

In the field of plant breeding, a breeder often wants to cluster
available genetic lines, characterized by a set of markers, to organize
the lines based on attributes of the population such as structure and
linkage disequilibrium [\citet{NC2010}]. They may also want to cluster
growing environments based on yield data of various lines to define a
target set of environments best suited to the line [\citet{CD1994}].
Clustering algorithms, where individuals or cases are assigned to
groups based on their similarity, is used. In many fields of science
where large amounts of data are being generated, clustering similar
cases or variables is often useful to organize the data. As in plant
breeding, cluster analysis is often used to answer specific questions.
Whether the research question is largely exploratory or inferential,
cluster analysis can contribute useful insight into the structure
hiding in a data set. Due to the underlying variation that is generally
unknown without genetic information, a major obstacle to cluster
analyses is estimating the number of clusters, which for genetic data
might be considered to be subpopulations. In fact, although current
clustering methods, such as $k$-means and hierarchical, are quite
useful, they do little to address the practical question of how many
clusters exist [\citet{FR2003}]. \citet{MC1985} tested
various procedures
for determining the number of clusters using classical hierarchical
methods, however, the simulated data was small and nonoverlapping, and
therefore not practical for genetic data. The methods for estimating
the number of clusters for $k$-means clustering has been reviewed
and include algorithmic, graphical and formulaic approaches [\citet
{S2006}]. Having insight into the number of clusters present for a
genetic marker data set can aid in understanding population structure.

Model-based clustering provides some help on choosing the
number of clusters by calculating some criterion based on the
population distribution assumptions. The most widely used model-based
clustering approach used in genetic studies is implemented in the
computer software STRUCTURE [\citet{PS2000}]. It decides the
number of
clusters by comparing variance-penalized log-likelihoods. STRUCTURE
has been cited in many research manuscripts. \citet{VE2007} applied
four separate rounds of STRUCTURE to Atlantic salmon (\emph{Salmo
solar}) genetic marker data and found that, although it seemed to work
well in clustering the genetic structure appropriately, the
computational time was intolerably long. \citet{HC2010} applied the
STRUCTURE model-based clustering to a large barley (\emph{Hordeum
vulgare} L.) data set consisting of 1816 individuals and 1416
variables (markers), wherein convergence did not occur after very
lengthy runs, finally requiring the use of another algorithm. In
addition to computational issues, STRUCTURE makes genetic assumptions
that are rarely met in breeding populations: (1) marker loci are
unlinked and in linkage equilibrium with one another within
populations, and (2) Hardy--Weinberg equilibrium within populations.
The first of these assumptions can be simply avoided by selection of
markers that are unlinked and in linkage equilibrium. In contrast, the
second assumption is rarely the case for plant breeding populations in
which selection plays a major role in population development. An
important result of these assumptions is that allele frequencies across
loci must be relatively similar, which is rarely the case for genetic
data.

For plant breeding, as in many other fields of science, the increasing
availability of data also results in high-dimensional data sets that
can be difficult computationally to cluster. The data that this paper
uses is binary data, presence or absence of a genetic marker, for each
unique line. There are commonly lots of missing values.
High-dimensionality issues related to cluster analyses were originally
described by \citet{B1961} as an exponential growth of hypervolume
as a
function of dimension. Clarifying this for clustering, \citet{M2009}
determined that in very high-dimensional space there is a
simplification of structure, demonstrating that the distances within
clusters become tighter, while between cluster distance expands, with
an increase in dimensionality. Though the research presented by
\citet
{M2009} makes a convincing argument to utilize all dimensions in
high-dimensional data sets, this is often not done due to the
computational burden. In addition, genetic data often times includes a
high frequency of nuisance variables that do not contribute to the
structure of the data. In order to overcome these possible issues, it
may be appropriate to implement cluster analyses on low-dimensional
projections such as the principal components (PCs) for some methods
[\citet{FR2002}]. 
\citet{HM2005} found that low-dimensional projections of such data sets,
where the number of dimensions $d \rightarrow\infty$ while the number
of observations $n$ is fixed, tend to lie at vertices of a regular
simplex, in agreement with \citet{M2009} and \citet{Ahn2007}.
(Note that
the supplementary material for this paper [\citet{Newetal13}] contains video of
higher-dimensional views of plant breeding data that also support the
claim that this simplified structure is present in these genetic data sets.)

HDLSS data can pose a challenge when applying principal component
analysis (PCA) because the covariance matrix is not of full rank. This
leads to a strong inconsistency in the lower eigenvectors, in which
case the added variation obscures the underlying structure of the
covariance matrix [\citet{JM2009}]. The first few eigenvectors are
consistent if there is a large difference in size of the eigenvalues
between these and the rest. 
Classic studies of dimension reduction and cluster analysis [e.g.,
\citet
{Chang1983}] caution against using PCA before clustering. The reason is
that PCA is finding directions of maximum variance in the data, which
does not always correspond to differences between clusters. This is
well known and explored further in several other papers, although it is
still mistakenly done. Projection pursuit, particularly with the holes
index [\citet{CBC93,SBH2012}], is a better approach for reducing
dimension before clustering. Ideally, clustering is done without
reducing the dimension, but some clustering methods that do not work
well for high-dimensional data, such as model-based clustering that
depends on estimating variance--covariance matrices, require a dimension
reduction step.

Advances in technology enable simulation of genetic data sets with
known cluster classifications. This application allows better testing
and evaluation of new algorithms on data sets with known properties.
Comparisons can also be made between simulated and empirical data sets
to gain insight into empirical data sets. The computer software GENOME
[\citet{LZ2007}], a coalescent-based whole genome simulator,
offers just
this by simulating sequences backward in time. Simulation of genetic
sequences is conditional on chosen parameters including, but not
limited to, population size, recombination rate and rates of migration
between subpopulations. The software is particularly fast so it has the
ability to generate a large number of data sets in a relatively short
period of time. Most importantly, setting the available parameters
enables the user to simulate data sets similar to empirical sets with
respect to the number of clusters and the level of structure present.
\citet{MC1985} evaluated different methods for determining the
number of
clusters, however, the simulated data was very limited in the number of
observations, number of dimensions and, most importantly for genetic
data, the lack of any cluster overlap.

The clustering methods that are currently available result in
distinctive outcomes that are often compared by the researcher on some
criterion and chosen accordingly. An approach that implements the array
of clustering methods available and chooses the method that minimizes
or maximizes a common criterion would be a useful approach that could
capitalize on the positives associated with specific methods. This
paper describes such an approach, that identifies the number of
clusters for genetic marker data that incorporates model-based,
$k$-means and hierarchical methods, and uses bootstrapping and cluster
criterion to help decide the number of clusters. The algorithm is
validated using GENOME simulated data and assessed on six empirical
data sets. Outcomes of the research include evaluation of an algorithm
to define the number of clusters using simulated data sets similar to
our empirical sets, comparison of simulated data sets to empirical data
sets, and development of graphical diagnostics to aid in the
determination of the number of clusters. We expect that these
contributions might be more generally applied to HDLSS data.

The paper is organized as follows. Section~\ref{algorithm} describes
the algorithm for choosing clusters. Section~\ref{data} describes the
simulated and empirical data sets used to validate the algorithm.
Section~\ref{results} describes the results. Supplementary material
[\citet{Newetal13}]
contains (1) the data sets, (2) R code for the analysis and (3) videos
of the data sets, and resulting clusters, shown in more than two
dimensions to better see the differences between clusters.

\section{Algorithm for choosing the number of clusters}\label{algorithm}

The algorithm to determine the number of clusters has four steps:
bootstrap sampling, clustering, calculation of a cluster validity
statistic, and the computation of a permutation test for significance.
Hubert's gamma statistic [\citet{HB2001}], available in the R package
\texttt{fpc} [\citet{Hennig2011}], is the cluster validity
statistic of
choice, chosen heuristically from many criteria within the algorithm
including the average distances within and between clusters and their
ratio, the within clusters sum of squares, the Dunn index and entropy.
Additionally, it is on a standard scale which makes comparison between
methods simpler and calculation across clustering methods trivial. For
consistency, matrices are denoted in bold typeface with the subscript
representing the number of rows and columns, respectively. Let $\blX
_{n\times p}$ ($n$~rows and $p$ columns) be the data set to be
clustered. In the genetic marker data, rows contain the lines and
columns the marker information. For the empirical sets, missing data
was imputed using the mean marker frequency for that marker, which is
%
%
\begin{figure}

\includegraphics{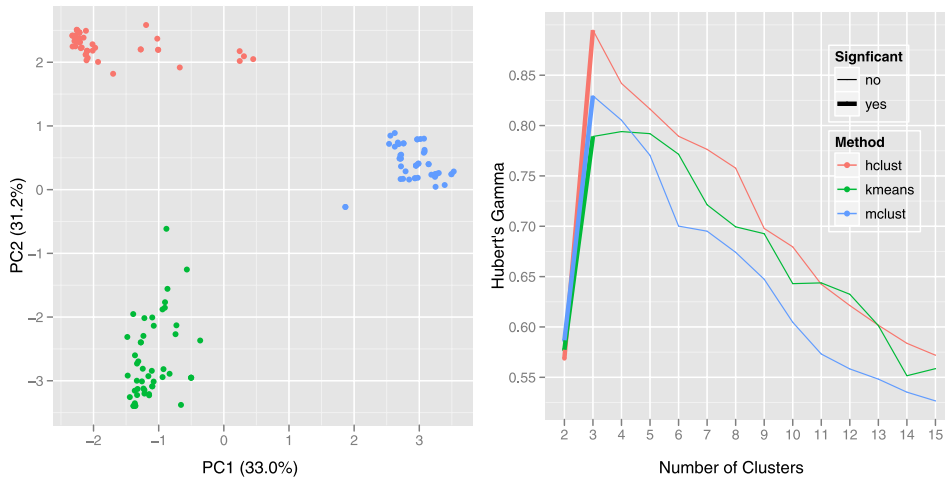}

\caption{(Left) principal component one (PC1) versus PC2 with percent
of the variation explained in parentheses for the example data set used
to show the steps of the proposed method. (Right) Hubert's gamma values
at each cluster number for the three methods of clustering on the
example simulated data set generated to have three clusters. Thick
lines represent significant ($p<0.01$) increases in Hubert's gamma for
pair-wise cluster numbers.}\label{fig1}
\end{figure}
common practice for genetic data. In addition, the steps are
graphically displayed using a $n=150$ by $p=100$ simulated data set
using STRUCTURE composed of three clusters of equal size with a
migration rate of 0.00001. The cluster structure is displayed using the
first two principal components, which for this data is shown in Figure~\ref{fig1}(left). The user sets a maximal number of clusters,
$k_{\mathrm{max}}$, based on prior knowledge of a maximum. The steps are then as follows:

\begin{enumerate}
\item\emph{Bootstrapping}: a number of bootstrap samples, $b$, are
drawn at random from the
rows of $\blX$ with replacement. The resulting matrix is denoted as
$\blX_{n\times p}^{*i}$ for $i=1,\ldots, b$.
\item\emph{Cluster analysis}: three methods of cluster analysis are
implemented for $1, 2,\ldots,k_{\mathrm{max}}$ clusters including model-based,
hierarchical and
$k$-means clustering.
\begin{enumerate}[(a)]
\item[(a)] Model-based (mclust) cluster analysis, available in the R
package \texttt{mclust} [\citet{FR2011}], is applied to principal
components $\blY_{n\times1}$, $\blY_{n\times2}$, $\blY_{n\times3},\ldots
,\blY_{n\times k_{\mathrm{max}}}$ where the number of clusters is set to
$k$. Thus, the
number of principal components is equal to the number of clusters. The
principal components were used only for the mclust method.
\item[(b)] Hierarchical (hclust) clustering is applied to the Manhattan
distance matrix $\blD_{n\times n}$ and cut at $k$ clusters. The
Manhattan distance was preferred to Euclidean distance, as it
represents the absolute distance between lines based on their binary
marker data.
\item[(c)] $k$-means (kmeans) clustering is applied to the bootstrap
sample $\blX_{n\times p}^{*i}$ with the
number of clusters set to $k$.
\end{enumerate}
\item\emph{Cluster validity}: for each $1, 2,\ldots,k_{\mathrm{max}}$
clusters, Hubert's gamma is calculated for model-based, hierarchical
and $k$-means
clustering on the Manhattan distance matrix. This results in three
Hubert's gamma statistics at each of $1, 2,\ldots,k_{\mathrm{max}}$ number of clusters.
\item\emph{Permutation test}: a paired permutation $t$-test is
computed for each consecutive
number of clusters across bootstrap samples for each method of clustering,
meaning between clusters 2:3, 3:$4,\ldots\,$, and $k-1$:$k_{\mathrm{max}}$. A linear
model is applied to
each pair with Hubert's gamma as the response and the cluster number as the
explanatory variable.
\item\emph{Choosing the number of clusters}: the clustering method
resulting in the highest
Hubert's gamma is used. The algorithm returns the lowest number of
clusters for
which Hubert's gamma is significantly greater than the number below it, but
not for the number above it. Results for the example data set are shown
in Figure~\ref{fig1}(right), with bold lines representing significant increases
in Hubert's gamma between
consecutive cluster pairs. For the example data set, all clustering
methods would
return three clusters; hierarchical clustering yielded the
highest Hubert's gamma, so it would be used.
\end{enumerate}

\section{Data}\label{data}

\subsection{Simulated}

In order to validate the proposed method, data sets were simulated
with varying numbers of clusters and degree of separation between
clusters. The coalescent whole genome simulator GENOME was used for all
simulations and was chosen because it was able to simulate data sets
covering the spectrum of variation in our empirical sets. The simulated
sets ranged in the number of clusters including 1, 2, 3, 4, 5, 6, 9 and
12 clusters. The level of separation between clusters was specified by
adjusting the migration rate per generation per individual, levels for
this parameter were 0.00005, 0.0001 and 0.00015. High levels of
migration resulted in less separated clusters, while low levels of
migration resulted in more separated clusters. The number of clusters
and migration rate were arranged as a factorial such that 100
simulations were tested at each cluster---migration rate combination.
All simulated sets included 200 observations and 400 markers, with each
cluster having equal numbers of observations. Because the number of
observations per simulation was fixed at 200, as the number of clusters
%
%
\begin{figure}

\includegraphics{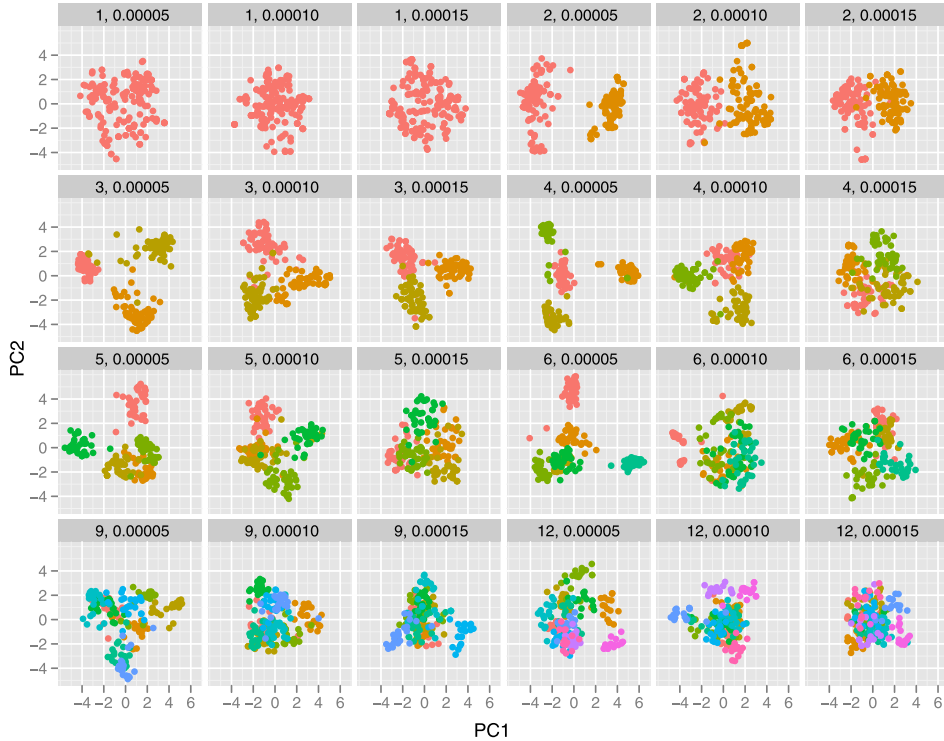}

\caption{PC1 versus PC2 for a randomly selected simulated data set for
each cluster---migration rate combination. Note that as the migration
rate increases, clusters are generally more overlapped. With more
clusters, the first two PCs are insufficient to capture the separation
of clusters, but it can still be seen that the clusters are further
apart with a lower migration rate.} \label{fig3}
\end{figure}
increased, the number of observations per cluster decreased. The
simulated data sets were HDLSS, which is generally the case for genetic
data sets. In order to gauge the variation in the simulated sets,
Figure~\ref{fig3} shows the first two principal components for one
random sample of each cluster---migration rate combination. Visually,
the simulated sets cover a wide spectrum of variability with respect to
the number of clusters and, most notably, the level of separation.

\subsection{Empirical}

Six empirical data sets were used from three small grain crops,
including three oat (\emph{Avena sativa} L.), two barley and one wheat
(\emph{Triticum aestivum} L.) data set. The first oat data set,
referred to as\break  \textsf{newell2010}, is a collection that includes
varieties, breeding lines and landraces of worldwide origin originally
used for analysis of population structure and linkage disequilibrium
[\citet{NC2010}]. The \textsf{newell2010} data set has 1205 observations
and 402 Diversity array technology (DArT) markers, which are binary,
with 5.1\% missing data. The second oat data set, referred to as \textsf
{tinker2009}, is also a set of varieties, breeding lines and varieties
of global origin that was used by \citet{TK2009} in the initial DArT
development work. The \textsf{tinker2009} data set consists of 198
observations and 1958 DArT markers with 21.6\% missing data. The third
oat data set, referred to as \textsf{asoro2011}, consists of 446 North
American elite lines scored for 1005 DArT markers with 5.8\% missing
data [\citet{AN2011}]. We note that there is some overlap between the
\textsf{newell2010} data set with both the \textsf{tinker2009} and
\textsf{asoro2011} data sets. This is because the \textsf{newell2010}
data set
combined data sets from independently assembled collections. Although
some observations are duplicated from the two sets in \textsf{newell2010},
all three data sets have different combinations of marker data, thus
they will cluster quite differently.

The first barley data set, referred to as \textsf{hamblin2010}, was
originally used to explore population structure and linkage
disequilibrium [\citet{HC2010}]. This set is the largest used in this
study and consists of 1816 observations from ten barley coordinated
agricultural project (CAP) participating breeding programs throughout
the US and scored for 1416 single nucleotide polymorphisms (SNPs), with
only 0.2\% missing data. Unlike the oat data sets, \textsf{hamblin2010}
has strong population structure, thus enabling testing of a wide
variety of cluster separation in the empirical sets. The second barley
data set, referred to as \textsf{zhang2009}, was originally used to assess
barley population structure and linkage disequilibrium [\citet{ZM2009}].
The data set is comprised of 169 lines consisting of mainly Canadian
cultivars and breeding lines scored on 971 DArT markers. The \textsf
{zhang2009} data set has about 2.6\% missing data. The last empirical
data set, referred to as \textsf{chao2010}, is a wheat data set also
originally used to explore population structure and linkage
disequilibrium [\citet{CD2010}]. The data set consists of 849 SNPs
scored on 478 spring and winter wheat cultivars from 17 breeding
programs across the US and Mexico. The \textsf{chao2010} data set contains
0.9\% missing data.

%
\begin{figure}

\includegraphics{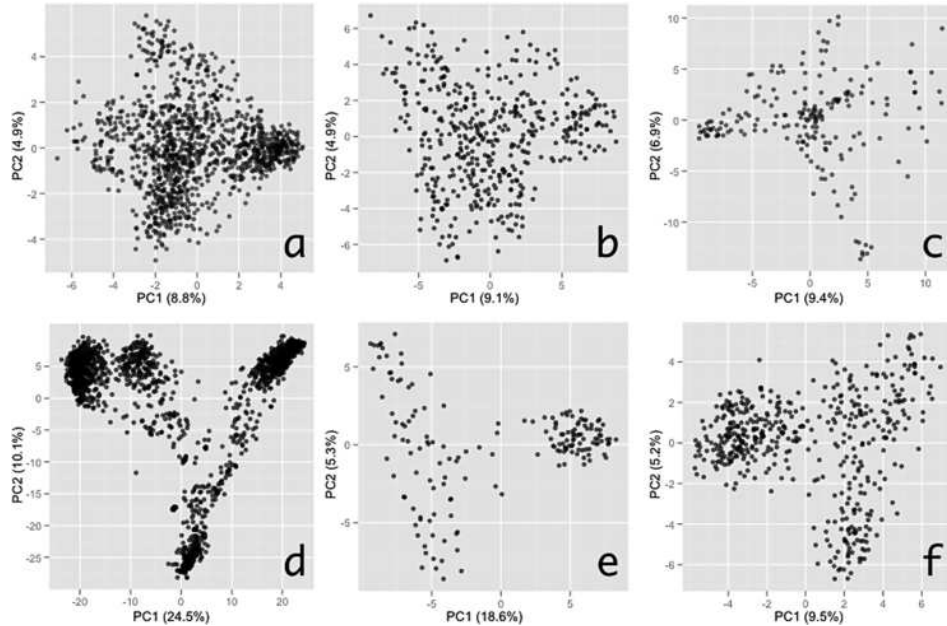}

\caption{The large amount of variation across the empirical sets by
visualization of the
first two principal components with amount of variation explained by
each axes in parentheses for \textup{(a)} \textsf{newell2010}, \textup
{(b)} \textsf{asoro2011},
\textup{(c)} \textsf{tinker2009}, \textup{(d)}
\textsf{hamblin2010}, \textup{(e)} \textsf{zhang2009}, and \textup{(f)}~\textsf{chao2010}.}
\label{fig4}
\end{figure}

%
%
\begin{figure}

\includegraphics{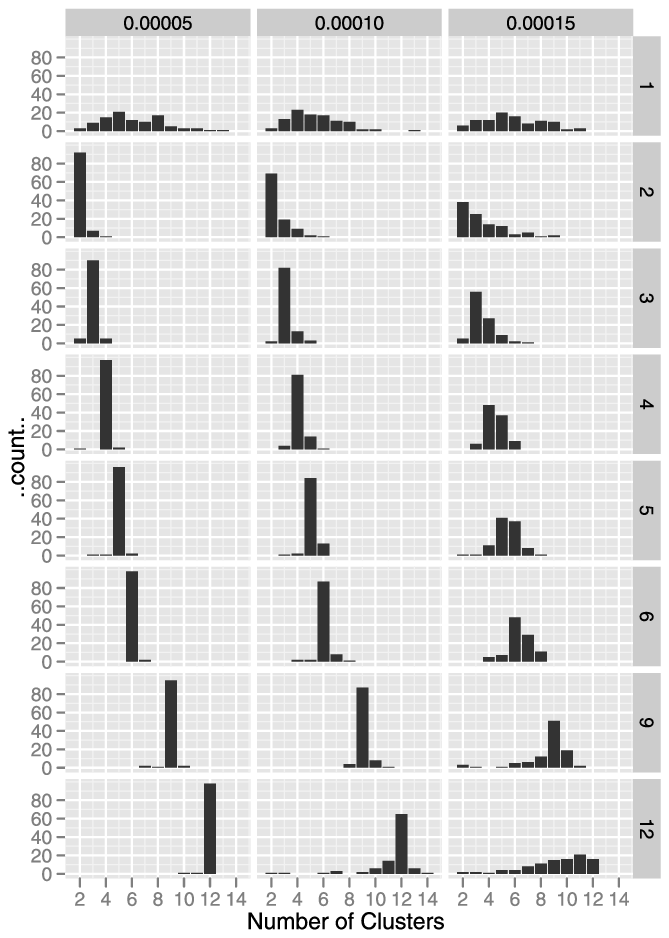}

\caption{Bar graph showing the resulting number of clusters
implementing the proposed algorithm for
100 simulations at each cluster---migration rate combination. The
numbers at the top and
right of each facet represent the migration rate and the true number of
clusters simulated,
respectively.}
\label{fig5}
\end{figure}

Taken together, the empirical data sets used in this study cover a
wide range of variation with respect to the level of separation between
clusters. The variation across empirical data sets can easily be seen
from their first two principal components (Figure~\ref{fig4}). The oat
data sets, \textsf{newell2010}, \textsf{asoro2011} and \textsf
{tinker2009}, have
relatively weak structure with less distinct clusters. In contrast, the
barley data sets, \textsf{hamblin2010} and \textsf{zhang2009}, and
wheat data
set, \textsf{chao2010}, show relatively strong structure, and clusters can
easily be seen in the principal component plots. These differences in
the level of structure across crops are most likely explained by the
breeding processes implemented for the specific crops. For example,
oats include hulled, naked, spring and winter types, but breeding
generally occurs across the major types and for the most part, lines
are usually spring, hulled types. In contrast, barley includes 2-row,
6-row, spring and winter types in all combinations in which it is
common practice to cross individuals within the same type but not
between types. This leads to the strong structure seen for the first
two principal components relative to the oat data sets. Similarly, the
%
%
\begin{table}
\tabcolsep=0pt
\caption{Summary of the six empirical data sets used in this study
including the assigned name, source of the original publication, crop,
origins of lines included, types of lines included, the dimensions
designated rows x columns, and the marker type designated as DArT or
SNP for Diversity Array Technology or single nucleotide polymorphism,
respectively}\label{tab1}
\begin{tabular*}{\tablewidth}{@{\extracolsep{\fill}}lccccc@{}}
\hline
\textbf{Source} &&  &\textbf{Line}&  &\textbf{Marker} \\
\textbf{(name)} & \textbf{Crop} & \textbf{Origins} & \textbf{types}& \textbf{Dimensions}
&\textbf{type}\\
\hline
\citet{NC2010}& Oat &World &Varieties, breeding & $1205\times 403$& DArT
\\
(\textsf{newell2010}) & & & lines, landraces& & \\
[2pt]
\citet{AN2011} & Oat &North & Elite cultivars& $446\times 1005$& DArT\\
(\textsf{asoro2011})&&American&&& \\
[2pt]
\citet{TK2009} & Oat &World& Varieties, breeding& $198\times 1958$&DArT\\
(\textsf{tinker2009}) & & & lines, landraces & & \\
[2pt]
\citet{HC2010} &Barley& US &Elite cultivars &$1816\times 1416$ &SNP\\
(\textsf{hamblin2010}) && & &&SNP\\
[2pt]
\citet{ZM2009} &Barley &Canada &Cultivars, & $169\times 971$&DArT \\
(\textsf{zhang2009}) & & & breeding lines && \\
[2pt]
\citet{CD2010} &Wheat &US, Mexico &Spring/winter & $478\times 219$ &
SNP\\
(\textsf{chao2010})&&& wheat cultivars & &\\
\hline
\end{tabular*}
\end{table}
first two principal components for the wheat data set separated spring
and winter types and further split spring types into two based on their
region of development. This indicates that for wheat, crossing does not
occur between spring and winter types and crossing most likely does not
occur across major regions within the spring types. The principal
component plots also allow comparison of the empirical and simulated
sets. Comparison of Figures~\ref{fig4} and \ref{fig5} demonstrate how
the low-dimensional
projections from PCA are quite similar between the simulated and
empirical sets and, more importantly, the simulated sets cover the
range of possibilities encountered in real data. A summary of the
empirical data sets used in this study is shown in Table~\ref{tab1}.

\section{Results}\label{results}

\subsection{Simulated data}

Results for 100 simulated data sets at each cluster---migration rate
combination are summarized in Table~\ref{tab2}. The mean estimated
number of clusters at the lowest migration rate was within 0.09 of the
true number of clusters across all combinations, excluding the case
%
%
\begin{table}
\caption{Summary of results for the simulated data sets including the
true number of clusters and migration rate simulated, mean estimated
number of clusters, true Hubert's gamma, Hubert's gamma, and the
proportion of times the correct number of clusters was chosen}
\label{tab2}
\begin{tabular*}{\tablewidth}{@{\extracolsep{\fill}}lcd{2.2}ccc@{}}
\hline
\textbf{True} & \textbf{Migration} & \multicolumn{1}{c}{\textbf{Mean}} & \textbf{True Hubert's}
& \textbf{Hubert's} & \textbf{Proportion} \\
\textbf{no} & \textbf{rate} & \multicolumn{1}{c}{\textbf{est no}} & \textbf{Gamma} & \textbf{Gamma} & \textbf{correct} \\
\hline
\hphantom{0}1 & 0.00005 & 6.09 & 1.00 & 0.48 & -- \\
& 0.00010 & 5.37 & 1.00 & 0.47 & -- \\
& 0.00015 & 5.79 & 1.00 & 0.48 & -- \\[3pt]
\hphantom{0}2 & 0.00005 & 2.09 & 0.69 & 0.73 & 0.92 \\
& 0.00010 & 2.47 & 0.49 & 0.55 & 0.69 \\
& 0.00015 & 3.46 & 0.36 & 0.48 & 0.38 \\[3pt]
\hphantom{0}3 & 0.00005 & 3.00 & 0.77 & 0.79 & 0.90 \\
& 0.00010 & 3.17 & 0.58 & 0.61 & 0.82 \\
& 0.00015 & 3.50 & 0.46 & 0.52 & 0.56 \\[3pt]
\hphantom{0}4 & 0.00005 & 4.00 & 0.78 & 0.79 & 0.97 \\
& 0.00010 & 4.12 & 0.59 & 0.62 & 0.81 \\
& 0.00015 & 4.49 & 0.46 & 0.51 & 0.48 \\[3pt]
\hphantom{0}5 & 0.00005 & 4.99 & 0.76 & 0.78 & 0.96 \\
& 0.00010 & 5.09 & 0.59 & 0.63 & 0.84 \\
& 0.00015 & 5.40 & 0.47 & 0.51 & 0.41 \\[3pt]
\hphantom{0}6 & 0.00005 & 6.02 & 0.75 & 0.77 & 0.98 \\
& 0.00010 & 6.04 & 0.58 & 0.62 & 0.87 \\
& 0.00015 & 6.34 & 0.46 & 0.51 & 0.48 \\[3pt]
\hphantom{0}9 & 0.00005 & 8.97 & 0.72 & 0.75 & 0.95 \\
& 0.00010 & 9.06 & 0.55 & 0.59 & 0.87 \\
& 0.00015 & 8.53 & 0.43 & 0.46 & 0.51 \\[3pt]
12 & 0.00005 & 11.97 & 0.68 & 0.71 & 0.98 \\
& 0.00010 & 11.36 & 0.51 & 0.55 & 0.65 \\
& 0.00015 & 9.20 & 0.41 & 0.43 & 0.16 \\\hline
\end{tabular*}
\end{table}
when one cluster was simulated. Hclust was the preferred method of
clustering, followed by kmeans and mclust which were chosen based on
the algorithm in 69\%, 28\% and 3\% of the simulations. As expected for
the largest migration rate, the mean estimated number of clusters
deviated the most from the true simulated number of clusters across all
combinations. Overall, the proportion of times the algorithm chose the
correct number of clusters ranged from 0.16 when 12 clusters were
simulated at the largest migration rate and 0.98 when six and 12
clusters were simulated at the lowest migration rate. In general, the
proportion of times the algorithm chose the correct number of clusters
decreased as the migration rate was increased. This was expected given
the fact that as the migration rate is increased, the clusters become
less distinct with more overlapping. These results are also shown in
Figure~\ref{fig5} at each cluster---migration rate combination. Because
the true classifications are known, a comparison between the true and
estimated Hubert's gamma across bootstrap samples can be made. Across
all simulations the true and estimated values of the Hubert's gamma
statistic decreased as the migration rate was increased. Likewise, in
all cases the estimated Hubert's gamma was larger than the true value;
this trend can also be seen in the mean estimated number of clusters
where this tends to overestimate the true number of clusters.

%
\begin{figure}

\includegraphics{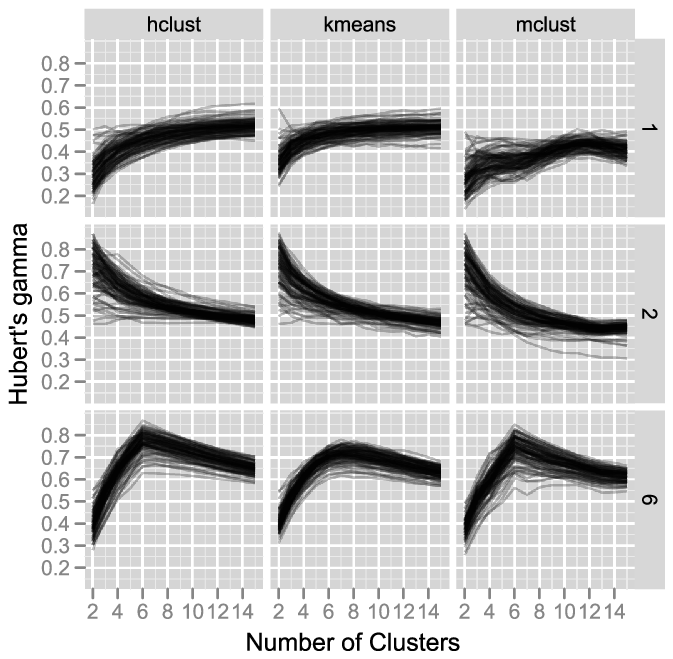}

\caption{Diagnostic plot of Hubert's gamma for varying numbers of
clusters to distinguish between the case when either one or two
clusters are present, illustrated on simulated data. One, two and six
cluster results are shown in the rows, and cluster method,
hierarchical, $k$-means and model-based clustering methods in the
columns. Of most importance is the different pattern between 1 and 2
clusters: in the case of just one cluster, Hubert's gamma increases
from 0.2, but in the case of two clusters there is a gradual decline in
Hubert's gamma from 0.7 with increasing number of clusters. For six
clusters, a distinctive peak occurs at 6.} \label{fig6}
\end{figure}

Like other methods of clustering, the algorithm does not directly have
the ability to detect the case when no structure exists, as is the case
when one cluster is simulated. The predicted number of clusters when
one cluster was simulated covered the range of possible values with no
definitive result across simulations (Figure~\ref{fig5}). For this
reason, it is important to have a diagnostic to determine when this is
in fact the case. The shape of the Hubert's gamma statistic relative to
the number of clusters can distinguish the case when no structure is
present. Figure~\ref{fig6} shows the shape of the Hubert's gamma
statistic for the three methods of clustering when one, two and six
clusters were simulated at the lowest migration rate for each
simulation. As depicted, when one cluster is simulated the shape of the
Hubert's gamma increases and levels off for hclust and kmeans with no
decrease in Hubert's gamma; this occurs in the opposite direction in
the case when two clusters were simulated. Likewise, for the case when
six clusters were simulated, the Hubert's gamma increased to a peak
followed by a decrease. Although these shapes can help distinguish the
case when one or two clusters are present, this can become quite
difficult when the migration rate is increased due to the fact that the
peak becomes less profound as the clusters become less distinct. In
addition to this diagnostic plot, the Hubert's gamma is positively
correlated with the proportion of times the correct number of clusters
was chosen with a value of 0.84 (Table~\ref{tab2}). Therefore, a~low
Hubert's gamma statistic for a data set gives an indication into the
confidence that the correct number of clusters was called. Thus far,
the results have been presented as if no prior information is known.
For genetic data sets this is rarely the case and can also be exploited
in choosing the final number of clusters.

\subsection{Empirical data}

\begin{figure}[b]
\vspace*{-3pt}
\includegraphics{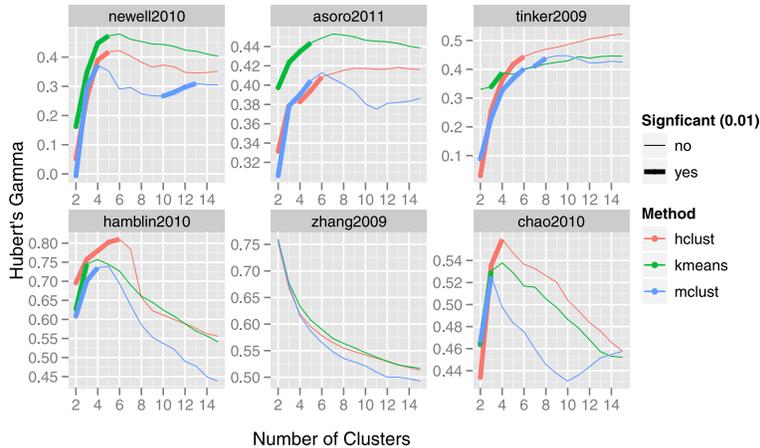}

\caption{Number of clusters versus the Hubert's gamma statistic for the
six empirical sets
in the study for 200 bootstrap samples. Colors refer to the three
clustering methods and
bold lines represent significant increase of Hubert's gamma for each
consecutive cluster
pair at $p<0.01$.}
\label{fig7}
\end{figure}

The empirical data sets imposed more variability with respect to the
degree of separation between clusters, number of lines per cluster and
the number of markers per data set. Table~\ref{tab3} summarizes the
algorithm results for the empirical sets used in this study for 50, 100
and 200 bootstrap samples. Results presented throughout will be for 200
bootstrap samples unless otherwise stated. The final numbers of
%
%
\begin{table}
\caption{Summary of results for the six empirical data sets in this
study including the final number of clusters, method and Hubert's gamma
shown in parentheses for 50, 100 and 200 bootstrap samples, and
previous results}\label{tab3}
\begin{tabular*}{\tablewidth}{@{\extracolsep{\fill}}lccccc@{}}
\hline
& \multicolumn{3}{c}{\textbf{No of clusters, method (Hubert's
Gamma)}} & \multirow{2}{36pt}[-7pt]{\centering{\textbf{Previous results}}} &
\multicolumn{1}{c@{}}{\multirow{2}{36pt}[-7pt]{\centering{\textbf{Previous method}}}}\\[-4pt]
& \multicolumn{3}{c}{\hrulefill} &&\\
\textbf{Data set} & \textbf{50} & \textbf{100} & \textbf{200} & &  \\
\hline
\textsf{newell2010} & 4, kmeans & 4, kmeans & 5, kmeans & 6 (0.438) &
mclust \\
& (0.481) & (0.481) & (0.526) & & on PCA \\[3pt]
\textsf{asoro2011} & 3, kmeans & 4, kmeans & 5, kmeans & 3 (0.434) &
kmeans \\
& (0.434) & (0.424) & (0.431) & & \\[3pt]
\textsf{tinker2009} & 1, -- & 1, -- & 1, -- & None & PCA and \\
& (1) & (1) & (1) & specified & hclust \\[3pt]
\textsf{hamblin2010} & 6, hclust & 6, hclust & 6, hclust & 7 (0.590) &
STRUCTURE \\
& (0.816) & (0.816) & (0.816) & & \\[3pt]
\textsf{zhang2009} & 2, kmeans & 2, kmeans & 2, kmeans & 2 (0.774) & PCA,
prior \\
& (0.786) & (0.786) & (0.786) & & knowledge \\[3pt]
\textsf{chao2010} & 3, kmeans & 4, hclust & 4, hclust & 9 (--) & STRUCTURE
\\
& (0.581) & (0.579) & (0.579) & & \\\hline
\end{tabular*} \vspace*{-3pt}
\end{table}
clusters for the six data sets ranged from one to six and are also
represented as the number of clusters versus the Hubert's gamma
statistic in Figure~\ref{fig7}. This plot is the diagnostic plot
presented in the simulation results. As shown, the starting value for
%
%
the Hubert's gamma statistic at two clusters covered a wide range
across the three clustering methods. zhang2009 has a
unique shape
that is characteristic of the case when two clusters are present.
\textsf{Newell2010}, \textsf{hamblin2010} and \textsf{chao2010} all
have distinct
peaks for all methods of clustering, indicating that greater than one
cluster is present. \textsf{Asoro2011} shows an increase in Hubert's gamma
for kmeans until about seven clusters, at which time it decreases, also
indicating that there is greater than one cluster. In contrast, \textsf
{tinker2009} is the only data set that is characteristic of the
situation in which only one cluster exists. If greater than one cluster
was true, the algorithm would identify six clusters using hclust. Due
to the fact that the Hubert's gamma using hclust does not show a peak
but a continuous increase, it is concluded that \textsf{tinker2009} has
only one cluster.

In order to assess the appropriate number of bootstrap samples
required for the empirical sets, the algorithm was applied using 50,
100 and 200 bootstrap samples (Table~\ref{tab3}). Results for two of
the data sets, \textsf{hamblin2010} and \textsf{zhang2009}, did not change
beyond 50 bootstrap samples, indicating that this was sufficient for
these data sets. The results for \textsf{chao2010} did not change beyond
100 boostrap samples, in which case this would be sufficient for this
data set. \textsf{Newell2010} required 200 bootstrap samples to reach
equilibrium with respect to the number of clusters; data is not shown
for 300. The results for \textsf{asoro2011} are unusual in the sense that
the number of clusters is still changing up to 200 bootstrap samples.
The algorithm was further tested for this data set using 300 and 400
bootstrap samples, where the number of clusters identified was six and
five, respectively. This outcome can be justified by the nature of the
data set, where the lines included are all North American elite oats
with a narrow genetic base. For a data set such as this it would be
concluded that the true number of clusters would be in the range of
five to six; in this case any prior information about the data set
would be helpful in a final decision. Interestingly, the number of
bootstrap samples required is negatively related to the Hubert's gamma
statistic for all of the data sets. \textsf{Asoro2011} requires the most
bootstrap samples and has the lowest Hubert's gamma, and \textsf
{hamblin2010} and \textsf{zhang2009} require the fewest number of bootstrap
samples and have the highest Hubert's gamma statistics. Application of
the results of the Hubert's gamma statistics at 50 bootstrap samples
can be used as an indicator for the number of bootstrap samples
required for a particular data set. For example, data sets with a
Hubert's gamma in the range of 0.786 to 0.816 only require 50 bootstrap
samples, those in the range of 0.581 require 100, those in the range of
0.481 require 200, and less than 0.434 require greater than 200
bootstrap samples, although, with a sample size of only six,
application to a greater number of empirical sets would be required to
solidify this claim. In summary, data sets resulting in larger Hubert's
gamma statistics require less bootstrap samples and, from the
simulation results, are more likely to determine the correct number of clusters.

Previous results for the six empirical sets are shown in Table~\ref{tab3} along with the method used for each result. As expected, the
number of clusters determined by the proposed algorithm differs in most
cases from previous results given the varying selection criteria across
methods. The previous method implemented for \textsf{newell2010}
identified six clusters using model-based cluster analysis implemented
on the first five principal components. In that study, the number of
clusters was based\vadjust{\goodbreak} largely on visual representation of principal
components, thus, it was largely user defined. In contrast, the
proposed algorithm defined five clusters using $k$-means
clustering. \citet{AN2011} identified three clusters for the
\textsf{asoro2011} data set, but also indicated that this number was chosen
based on the research objectives for that study; six clusters were
initially identified. Previous results for \textsf{tinker2009} did not
necessarily identify a certain set number of clusters but used
clustering more as a general guide to study the diversity of lines. The
lines used in \citet{TK2009} were initially chosen to represent the
diversity of oat on a worldwide scale; this can be seen in the first
two PCs where lines tend to spread from a point resembling a bull's-eye
(Figure~\ref{fig4}). The algorithm identified only one cluster for this
data set, which does conform to how the data was initially chosen.
Similar results were found for the \textsf{hamblin2010} data set by
implementation of the proposed algorithm and STRUCTURE [\citet{HC2010}],
where six and seven clusters were identified, respectively. Results
presented by \citet{ZM2009} were the same for the proposed algorithm,
with identification of two clusters. Last, the results for the \textsf
{chao2010} were largely different, with four and nine clusters
identified for this algorithm and \citet{CD2010}, respectively.
The four
clusters identified by the algorithm respond to the group of winter and
spring lines split into three groups. Overall, the proposed algorithm
identifies a similar number of clusters to previous methods but is
different considering the criterion for which the number of clusters is chosen.

%
\begin{figure}[b]

\includegraphics{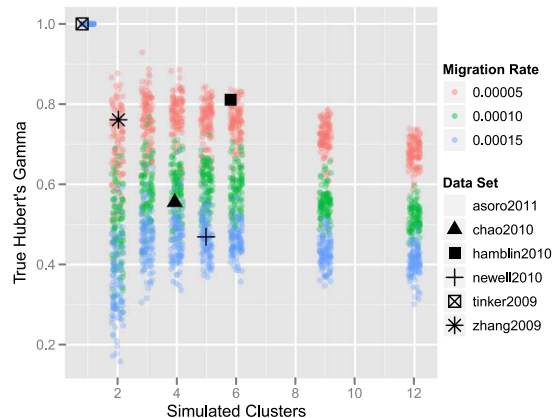}

\caption{True Hubert's gamma values for all simulated data sets colored
by migration rate overlaid with empirically determined Hubert's gamma
values for the empirical data sets. This plot gives some suggestions
for the migration rate observed with the empirical data: low for \textsf
{zhang2009} and \textsf{hamblin2010}, high for \textsf{asoro2011} and
\textsf{newell2010}, and medium for \textsf{chao2010}.} \label{fig8}
\end{figure}

In order to gain insight into where the empirical sets fall with
respect to the simulated sets, Hubert's gamma statistics for each are
shown simultaneously in Figure~\ref{fig8}. The variation in the true
Hubert's gamma for the simulated data sets at each cluster---migration
rate combination covers a range of about 0.8, in which case a lower
migration rate has a higher Hubert's gamma. \textsf{Zhang2009} and
\textsf{hamblin2010} fall within the range of the lowest migration rate
at two
and six clusters, respectively. \textsf{Chao2010} falls within the range
of the middle migration rate, 0.0001, with four clusters. \textsf
{Newell2010} and \textsf{Asoro2011} fall within the range of the largest
migration rate of 0.00015, both at five clusters. Both the \textsf
{Newell2010} and \textsf{Asoro2011} data sets, in addition to falling
within the range of the largest migration rate, also have the smallest
Hubert's gamma statistics. Last, \textsf{tinker2009}, having only one
cluster, has a Hubert's gamma statistic of one. These comparisons can
provide some information into the confidence of the correct number of
clusters for the empirical sets. Empirical sets that fall within the
range of the smallest and largest migration rates would have relatively
more and less confidence, respectively.

\section{Conclusion}

This paper has proposed an algorithm that provides assistance in
choosing the number of clusters and the clustering algorithm for HDLSS
data. The algorithm uses bootstrap samples to quantify the cluster
variation and permutation tests on Hubert's gamma statistics to test
for significance of the chosen number of clusters. Validation of the
algorithm on HDLSS data simulated by GENOME with varying numbers of
clusters and level of separation indicates that the algorithm operates
well on data of this sort. As clusters get more overlapped, if the
migration rate is large, the accuracy in estimating the correct number
of clusters declines. For the case when no cluster structure is present
in a data set, a diagnostic plot of the change in Hubert's gamma across
varying numbers of clusters can be used to indicate the lack of clusters.

The results from this algorithm on six empirical data sets vary
slightly from the reported number of clusters in previous studies, but
are not wildly different. The empirical data sets vary less uniformly
than the simulated data sets, which might be expected. In most cases,
the change in Hubert's gamma across the number of clusters in the
simulated data resulted in significant peaks at the true simulated
number of clusters. The three clustering methods did result in largely
different Hubert's gamma statistics, with no one method being better
than the others on all data sets, demonstrating the importance of
including multiple clustering methods in the algorithm. However, it
should be pointed out that mclust was the preferred clustering method
in only 3\% of the simulated data sets and was never the preferred
clustering method in the empirical data sets. Previous research has
highly recommended against clustering the principal components
[\citet
{HA1985}] and termed it ``tandem clustering.'' Although the proposed
algorithm does in fact implement mclust on the principal components, it
rarely is actually selected as the best method.

In agreement with two previous studies [\citet{HM2005,M2009}], all of
the empirical sets, and the simulated data, exhibit a simplex shape in
the first few PCs. The different clusters form the vertices of the
simplex. A~comparison of the empirical to simulated sets illustrates
that the Hubert's gamma statistics of the empirical sets are within the
range of values observed for the simulated sets. This, along with the
visualization of the PCs, supports a conclusion that the GENOME
software is able to adequately simulate data sets that match well with
the empirical sets. By plotting the Hubert's gamma of the empirical
data sets in comparison to those of the simulated data sets for
different migration rates, a~reasonable sense of the migration rate
observed by the empirical data sets can be determined.

Finally, we expect the cluster selection algorithm might be applicable
to other HDLSS data. For other types of problems, where the data is not
binary as is for the genetic data used here, comparison data might be
simulated from a Gaussian mixture distribution for validation purposes.

\section*{Acknowledgments}

We would like to thank Eduard Akhunov, Franco Asoro, Francois Belzile
and Nick Tinker for providing empirical data sets for testing.

\begin{supplement}
\sname{Supplement A}
\stitle{Videos of High-Dimensional Views of Empirical and Simulated Data}
\slink[doi]{10.1214/13-AOAS671SUPPA} 
\sdatatype{.pdf}
\sfilename{aoas671\_suppa.pdf}
\sdescription{Video footage of tours of the empirical and simulated
data sets.}
\end{supplement}

\begin{supplement}
\sname{Supplement B}
\stitle{Empirical and Simulated Data\\}
\slink[doi,text={10.1214/13-AOAS671SUPPB}]{10.1214/13-AOAS671SUPPB} 
\sdatatype{.zip}
\sfilename{aoas671\_suppb.zip}
\sdescription{Data sets used in this paper.}
\end{supplement}

\begin{supplement}
\sname{Supplement C}
\stitle{R Code}
\slink[doi]{10.1214/13-AOAS671SUPPC} 
\sdatatype{.zip}
\sfilename{aoas671\_suppc.zip}
\sdescription{Software used to make the calculations for this paper.}
\end{supplement}

%

\printaddresses

\end{document}